\documentclass[11pt]{article}
\usepackage[pctex32]{graphics}

\def\nn{\nonumber} \def\bd{\begin{document}} \def\ed{\end{document}}
\def\ds{\documentstyle} \let\fr=\frac \let\bl=\bigl \let\br=\bigr
\let\Br=\Bigr \let\Bl=\Bigl
\let\bm=\bibitem
\let\na=\nabla
\let\pa=\partial \let\ov=\overline
\newcommand{\be}{\begin{equation}}
\newcommand{\ee}{\end{equation}}
\def\ba{\begin{array}}
\def\ea{\end{array}}
\def\ft#1#2{{\textstyle{\frac{\scriptstyle #1}{\scriptstyle #2} } }}
\def\fft#1#2{{\frac{#1}{#2}}}
\def\del{\partial}
\def\vp{\varphi}
\def\sst#1{{\scriptscriptstyle #1}}
\def\oneone{\rlap 1\mkern4mu{\rm l}}
\def\td{\tilde}
\def\wtd{\widetilde}
\def\ie{{\it i.e.\ }}
\def\dalemb#1#2{{\vbox{\hrule height .#2pt
        \hbox{\vrule width.#2pt height#1pt \kern#1pt
                \vrule width.#2pt}
        \hrule height.#2pt}}}
\def\square{\mathord{\dalemb{6.8}{7}\hbox{\hskip1pt}}}
\newcommand{\ho}[1]{$\, ^{#1}$}
\newcommand{\hoch}[1]{$\, ^{#1}$}
\newcommand{\bea}{\setlength\arraycolsep{2pt} \begin{eqnarray}}
\newcommand{\eea}{\end{eqnarray}}
\newcommand{\ra}{\rightarrow}
\newcommand{\lra}{\longrightarrow}
\newcommand{\Lra}{\Leftrightarrow}
\newcommand{\bp}{\tilde \beta^\prime}
\newcommand{\tr}{{\rm tr} }
\newcommand{\Tr}{{\rm Tr} }
\def\0{{\sst{(0)}}}
\def\1{{\sst{(1)}}}
\def\2{{\sst{(2)}}}
\def\3{{\sst{(3)}}}
\def\4{{\sst{(4)}}}
\def\5{{\sst{(5)}}}
\def\6{{\sst{(6)}}}
\def\7{{\sst{(7)}}}
\def\8{{\sst{(8)}}}

\begin{document}

\vspace{5mm}
\begin{center}
{\Large \bf Chiral gravitational waves from $z=2$
Ho\v{r}ava-Lifshitz gravity } \vspace{12mm}

{\large   Yun Soo Myung \footnote{e-mail
 address: ysmyung@inje.ac.kr}}
 \\
\vspace{10mm} {\em Institute of Basic Science and School of
Computer Aided Science \\ Inje University, Gimhae 621-749, Korea}
\end{center}

\begin{center}

\underline{Abstract}
\end{center}

  We construct the chiral gravitational waves from the $z=2$ Ho\v{r}ava-Lifshitz gravity with
  gravitational Chern-Simons term in the de Sitter and Minkowski backgrounds.
  These gravitational waves which show a feature  of  the Ho\v{r}ava-Lifshitz gravity
  may be related to  the generalized uncertainty principle. In addition,
  we find the classical and quantum IR-UV transition rules in the $z=2$ Ho\v{r}ava-Lifshitz gravity.\vspace{15pt}

\thispagestyle{empty}

%\pagebreak
%\voffset=0pt
%\setcounter{page}{1}

%\tableofcontents

%\addtocontents{toc}{\protect\setcounter{tocdepth}{2}}

%%%%%%%%%%%%%%%%%%%%%%%%%%%%%%%%%%%%%%%%

\newpage
\section{Introduction}
Recently Ho\v{r}ava has proposed a renormalizable theory of gravity
at a Lifshitz point~\cite{ho1},  which  may be regarded as a UV
complete candidate for general relativity. At short distances the
theory of $z=3$ Ho\v{r}ava-Lifshitz (HL) gravity describes
interacting nonrelativistic gravitons and is supposed to be
power-counting renormalizable in 3+1 dimensions. The equations of
motion were derived for $z=3$ HL gravity~\cite{KK,LMP}. Its
cosmological implication first appeared in~\cite{KK}, while its
black hole solution was  found in asymptotically anti-de Sitter
spacetimes~\cite{LMP} and black hole in asymptotically flat
spacetimes~\cite{KS}.

Even though the  $z=3$ HL gravity was proposed to be a
power-counting renormalizable theory in 3+1 dimensions, there are
many fundamental issues to be clarified. Renormalizability beyond
power-counting has not yet been proven and the renormalization group
(RG) flow of various coupling constants has not been studied.
Especially, the recovery of general relativity depends critically on
the assumption that the parameter $\lambda$ flows to 1 in the IR
limit. Since no one insists that the $z=3$ HL gravity is the final,
one may  either improve it  or modify it. In this sense,  we may
consider the $z=2$ HL gravity as an alternative because of the
difficulty in working with $z=3$ HL gravity.

One application of the $z=2$ HL gravity in 2+1 dimensions, as a
candidate membrane world-volume theory was discussed in~\cite{ho2}.
In this work, we wish to consider  the $z=2$ HL gravity in 3+1
dimensions as a candidate for the quantum theory  of  general
relativity. The reason is as follows. It was shown that the
renormalized Wheeler-DeWitt equation possesses a solution with a
$z=2$ Lifshitz point, but no other $z>2$ solutions to leading order
of strong coupling expansion~\cite{Sak}. This indicates that the
quantum Einstein gravity has a $z=2$ Lifshitz point, but no other
higher Lifshitz points. Adding the gravitational Chern-Simons (gCS)
term to the $z=2$ HL gravity leads to the fact that  their
conclusion remains unchanged.   The author in~\cite{Nish} has
obtained that the Ricci flow in ``$d$" spatial dimensions is the
holographic RG flow to the ($d+1$)-dimensional $z=2$ HL gravity with
$\lambda=1/2$. Also, there are some results, supporting that
perturbative corrections to $z=2$ Lifishitz point are promising.
These include the quantum Lifshitz model in $d+1$
dimensions~\cite{OR}, Lifshitz-type scalar field theory in $d=4$ and
$d=10$~\cite{IRS}, and Lifshitz-type gauge theory~\cite{Kaw}.

Concerning cosmological implications of the  $z=3$ HL gravity, it
has provided a new mechanism of generating scale-invariant cosmological
perturbations~\footnote{In this case, the scaling factor takes the
form  $a(t)\sim t^{p}(p>1/3)$. It may be regarded as an alternative
to inflation. However, there are a lot of questions. An urgent
question is how one does solve the flatness problem without
inflation~\cite{CQ}.}  \cite{Mukos} and regular bounce solutions
 in the early universe~\cite{Cal,Brand}.
Importantly, the authors in~\cite{TS} have shown that the chiral
primordial gravitational waves are generated from the $z=3$ HL
gravity when working the pure de Sitter cosmological background.
These circularly polarized modes are generated only when the Cotton
tensor $C_{ij}$ is present, making parity violation. Since these
modes are composed of higher order spatial  derivatives,  it may be
likely observed if these modes were really present in the very early
universe\footnote{ Concerning the observability of chiral
gravitational waves, the primordial gravitational waves should be
created at high energy in order to be observable. Once they are
created, their amplitudes will conserve until the final horizon
crossing.  On the other hand, the important point is the initial
condition. In the presence of the parity violating term, the initial
quantum state of gravitational waves has chirality. This is the
reason why one has obtained chiral primordial gravitational waves in
the parity violating theory~\cite{TS}.  It does not depend on the
number of derivatives. Thus, it is conjectured that there is no
essential difference between $z=3$ HL gravity and $z=2$ HL gravity
with gCS term.} . Let us study what happens in the $z=2$ HL gravity.
Aside from the fact that the scaling-invariant spectrum is a
perturbative feature of the $z=3$ HL gravity, the remaining two
could be achieved when using the $z=2$ HL gravity. A matter bounce
solution can be derived from the $z=2$ HL gravity with a matter,
since there is no essential difference in Friedmann equations
between $z=2$ and $z=3$ HL gravities. This is so because the Cotton
tensor from the gCS term did not contribute to the
Friedmann-Robertson-Walker (FRW) universe based on the isotropy and
homogeneity, but it contributes to the Mixmaster universe (Bianchi
IX) based on the anisotropy and homogeneity~\cite{mksp,Bakas}.
Hence, it is quite interesting to see whether chiral gravitational
waves without ghost instability can be generated from the $z=2$ HL
gravity.

On the other hand,  one of main ingredients for studying quantum
gravity is the generalized uncertainty principle (GUP), which has
been argued from various approaches to quantum gravity and black
hole physics~\cite{gup}. We note that the GUP is in the heart of the
quantum gravity phenomenology. Certain effects of quantum gravity
are universal and thus, influence almost any system with a
well-defined Hamiltonian~\cite{DV}.  It seems that the GUP may be
 related to black holes found in the deformed HL gravity~\cite{Myung,Myungent}. Furthermore, it was shown that the
GUP-corrected tensor propagator takes similar form as that derived
from the $z=3$ HL gravity~\cite{Myung3,Myung4}.

We wish to mention why the connection between $z=2$ HL gravity and
GUP will be  important to understand the quantum aspects of $z=2$ HL
garvity. The GUP usually satisfies the modified Heisenberg algebra $
[x_i,p_j]=i\hbar\Big(\delta_{ij}+\beta p^2 \delta_{ij}+2\beta
p_ip_j\Big)$,  where $p_i$ is considered as the momentum at high
energies.  Thus, it can be interpreted to be the UV-commutation
relation.  On the other hand, introducing IR-canonical variable
$p_{0i}$ with $x_i=x_{0i}$  via the replacement $p_i\to
p_{0i}\Big(1+\beta p_0^2\Big)$, these variables satisfy canonical
(IR) commutation relation $ [x_{0i},p_{0j}]=i\hbar\delta_{ij}.$ Here
$p_{0i}$ is considered as the momentum at low energies.  It is easy
to show that the UV-commutation  is satisfied to $\beta$-order when
using the IR-commutation. Hence, the replacement could be used as an
``important low-energy window" to investigate quantum gravity
phenomenology up to $\beta$-order. Similarly, if the UV-tensor
propagator of $z=2$ HL gravity which carries an information on the
renormalizability could be obtained from the IR-UV transition of
$p^2\to p^2(1+2p^2/\omega)$, one insists that the GUP can be
realized in the $z=2$ HL gravity partly.  We note that even though
the GUP is in the heart of the quantum gravity phenomenology, it
reveals  a part of quantum gravity effects but not whole of quantum
gravity effects.

In this work, we wish to construct chiral gravitational waves from
the $z=2$ HL gravity~\cite{ho2} with the gCS term in the de Sitter
and Minkowski backgrounds. Also we suggest that these gravitational
waves may be related to the GUP.

 \section{$z=2$ HL gravity with gCS term}
Introducing the ADM formalism where the metric is parameterized
%%%
\be ds_{ADM}^2= - N^2  dt^2 + g_{ij} \Big(dx^i - N^i dt\Big)
\Big(dx^j - N^j dt\Big)\,, \ee
%%%%
the Einstein-Hilbert action can be expressed as
%%%
\be \label{Eins} S_{EH} = \fft{1}{16\pi G} \int d^4x \sqrt{g} N
\Big[K_{ij} K^{ij} - K^2 + R - 2\Lambda\Big], \ee
%%%%
where $G$ is Newton's constant and extrinsic curvature $K_{ij}$
takes the form
%%%
\be K_{ij} = \fft{1}{2N} \Big(\dot g_{ij} - \nabla_i N_j -
\nabla_j N_i\Big)\,. \ee
%%%%
Here, a dot denotes a derivative with respect to $t$. The action of
$z=2$ HL gravity is given by~\cite{ho2} \be S_{HL}=\int dtd^3x {\cal
L}_{z=2}=\int dtd^3x \Big[{\cal L}^\lambda_K + {\cal
L}^\lambda_V\Big], \label{action1} \ee where the kinetic Lagrangian
is given by \be \label{kineticl}{\cal L}^\lambda_K
=\frac{2}{\kappa^2}\sqrt{g} N K_{ij}{\cal G}^{ijkl}K_{kl}= \sqrt{g}
N \frac{2}{\kappa^2}\Bigg(K_{ij}K^{ij}-\lambda K^2\Bigg), \ee with
the DeWitt metric
 \be {\cal G}^{ijkl}=
\frac{1}{2}\Big(g^{ik}g^{jl}-g^{il}g^{jk}\Big)-\lambda
g^{ij}g^{kl} \ee
 and its inverse metric
 \be {\cal
G}_{ijkl}=\frac{1}{2}\Big(g_{ik}g_{jl}-g_{il}g_{jk}\Big)-\frac{\lambda}{3\lambda-1}g_{ij}g_{kl}.\ee
The potential Lagrangian is determined by the detailed balance
condition as \bea &&{\cal L}^\lambda_V=-\frac{\kappa^2}{2}\sqrt{g}N
E^{ij}{\cal G}_{ijkl}E^{kl} \nn \\
&&= \sqrt{g}N \frac{\kappa^2\mu^2(-\Lambda_W)}{8(3\lambda-1)}
\Bigg(R-3\Lambda_W-\frac{4\lambda-1}{4\Lambda_W}R^2
+\frac{(3\lambda-1)R^2_{ij}}{\Lambda_W} \Bigg).\label{action2}
\eea Explicitly, $E_{ij}$ could be derived  from the Euclidean 3D
gravity \be E^{ij}=\frac{1}{\sqrt{g}} \frac{\delta W_{3D}}{\delta
g_{ij}} \ee with \be W_{3D}=- \mu \int d^3x
\sqrt{g}(R-2\Lambda_W). \ee

In order to generate  chiral gravitational modes, it is necessary to
introduce  the gravitational Chern-Simon (gCS) term~\cite{DJT} \be
S_{gCS}=\alpha\int dt d^3 x N\sqrt{g}{\cal L}_{gCS}=\alpha \int dt
d^3 x \sqrt{g}N \epsilon^{ikl}\Big(\Gamma^m_{il}\partial_j
\Gamma^l_{km}+\frac{2}{3} \Gamma^n_{il} \Gamma^l_{jm} \Gamma^m_{kn}
\Big) \ee whose variation with respect to $g_{ij}$ leads to the
Cotton tensor $C^{ij}$ \be \alpha C^{ij}=\alpha
\epsilon^{ik\ell}\nabla_k\left(R^j{}_\ell
-\frac14R\delta_\ell^j\right).\label{def.K.C} \ee Here
$\epsilon^{ikl}$ is a tensor and $\alpha$ is an arbitrary parameter
with scaling dimension $[\alpha]=2$.

In the IR limit,  comparing $S_{HL}$ with Eq.(\ref{Eins}) of
general relativity, the speed of light, Newton's constant and the
cosmological constant are given by
%%%%
\be c=\fft{\kappa^2\mu}{4}
\sqrt{\fft{\Lambda_W}{1-3\lambda}}\,,\qquad
G=\fft{\kappa^2}{32\pi\,c}\,,\qquad \Lambda_{\rm cc}=\ft32
\Lambda_W\,.\label{cg} \ee  Considering the $z=2$ HL gravity with
$[t]=-2,[x^i]=-1~([{\cal L}^\lambda_K]=[{\cal
L}^\lambda_V]=[W_{gCS}]=5)$, UV scaling dimensions are given by \be
[N]=0,~~[\mu]=1,~~[\kappa^2]=-1,~~[\Lambda_W]=2,~~[c]=1. \ee

 In the case of $\lambda=1(\Lambda_W<0)$, the $z=2$ HL
potential Lagrangian  with gCS term takes the form \be {\cal
L}_V^{\lambda=1}=
\sqrt{g}N\frac{2c^2}{\kappa^2}\Bigg(R-3\Lambda_W-\frac{3}{4\Lambda_W}R^2
+\frac{2}{\Lambda_W}R_{ij}^2+2\sqrt{-\frac{2}{\Lambda_W}}{\cal
L}_{gCS}
 \Bigg),\label{action3} \ee where
we recover the general relativity  with cosmological constant in the
limit of $\Lambda_W \to \infty$. This Lagrangian is useful to study
cosmological implications. It is obvious that ${\cal
L}_V^{\lambda=1}$ does not satisfy the detailed balance condition
because  the last term is present. Here,  we choose  \be
\alpha=\frac{4c^2}{\kappa^2}\sqrt{-\frac{2}{\Lambda_W}}=\frac{\kappa^2
\mu^2}{4}\sqrt{-\frac{\Lambda_W}{2}}=\mu c, \ee to ensure that
ghost-free chiral gravitational waves are propagating in the pure de
Sitter background. We could not obtain ghost-free chiral
gravitational waves unless $\alpha=\mu c$.

We would like to mention that the IR vacuum of this theory is
anti-de Sitter (AdS$_4$) spacetimes. Hence, it is interesting to
take a limit of the theory, which may lead to  a Minkowski vacuum in
the IR sector. To this end, one may deform the theory by adding
``$\mu^3R$" $(\tilde{{\cal L}}^\lambda_V={\cal
L}^\lambda_V+\sqrt{g}N \mu^3R)$ and then, taking  the $\Lambda_W \to
0$ limit~\cite{KS}. We call this the ``deformed $z=2$ HL gravity".
This does not alter the UV property of the theory, while it changes
the IR property. That is, there exists a Minkowski vacuum, instead
of an AdS vacuum. In the IR limit, the speed of light and Newton's
constant are determined  by
%%%%
\be c^2=\fft{\kappa^2\mu^3}{2},~
G=\fft{\kappa^2}{32\pi\,c}.\label{kh} \ee For $\lambda=1$, the
deformed $z=2$ HL gravity with gCS term takes the form \be
\label{action4} \tilde{{\cal L}}^{\lambda=1}_V=\sqrt{g}N
\frac{2c^2}{\kappa^2}\Bigg(R
+\frac{3}{4\omega}R^2-\frac{2}{\omega}R^2_{ij}+
2\sqrt{\frac{2}{\omega}}{\cal L}_{gCS} \Bigg),\ee where an important
parameter $\omega$~\cite{KS} and $\alpha$ are given by \be
\omega=\frac{16\mu}{\kappa^2},~~\alpha=\mu
c=2\sqrt{\frac{2}{\omega}}\ee with scaling dimension $[\omega]=2$.
We note that a choice of $\alpha=\mu c$ is essential to find
ghost-free chiral gravitational waves propagating  on the Minkowski
background.  In the limit of $\omega \to \infty$,
we recover the general relativity. Comparing ${\cal
L}^{\lambda=1}_V$ with $\tilde{{\cal L}}^{\lambda=1}_V$, these become
the same form when replacing $-1/\Lambda_W$ by $1/\omega$ up to
cosmological constant $-3\Lambda_W$.

\section{Cosmological implications}

First of all, we look for cosmological equation to $z=2$  HL gravity
with gCS  term by introducing the FRW metric \be \label{uvhamil1}
ds^2_{FRW}=-c^2dt^2+a(t)^2\Bigg[\frac{dr^2}{1-\bar{k}r^2}+r^2(d\theta^2+\sin^2\theta
d\phi^2)\Bigg] \ee
 where $\bar{k}=1,0,-1$ correspond to a closed, flat, and open
 universe, respectively.  For vacuum solution without matter
 ($p=\rho=0$), the first Friedmann equation takes the form~\cite{LMP}
 \be \label{HLF}
H^2=\frac{\Lambda_W}{2}-\frac{\bar{k}}{a^2}
\Bigg[1-\frac{1}{2\Lambda_W} \frac{\bar{k}}{a^2}\Bigg], \ee where
scaling dimensions are  $[\bar{k}]=2$ and $[H^2]=2$ with
$H=\frac{\dot{a}}{ca}$. Here,  we observe a replacement for  the
IR-UV transition\be \label{IRUVC1}\frac{\bar{k}}{a^2} \to
\frac{\bar{k}}{a^2}\Bigg(1-\frac{1}{2\Lambda_W}
\frac{\bar{k}}{a^2}\Bigg). \ee  However, we note that for the
$\bar{k}=0$ case, there is no contribution from higher order
curvature terms $R^2$ and $R^2_{ij}$. Also, there is no contribution
from the Cotton tensor  originated at gCS term because the FRW
metric (\ref{uvhamil1}) is based on the isotropy and homogeneity.
Thus,  we obtain the same Friedmann equation even if one considers
the $z=3$ HL gravity. In the presence of a matter, there were bounce
solutions to Friedmann equations~\cite{Brand,CSa,Park}. Also,
classical solutions to the IR limit of Ho\v{r}ava-Lifshitz gravity
can mimic general relativity plus cold dark matter~\cite{Muko}.

In the case of deformed $z=2$ HL gravity with gCS term, the
corresponding Friedmann equation leads to~\cite{KS} \be \label{DHLF}
H^2=-\frac{\bar{k}}{a^2}\Bigg[1+\frac{1}{2\omega}\frac{\bar{k}}{a^2}\Bigg],
\ee where we find  a replacement for the  IR-UV transition \be
\label{IRUVC2}\frac{\bar{k}}{a^2} \to
\frac{\bar{k}}{a^2}\Bigg(1+\frac{1}{2\omega}\frac{\bar{k}}{a^2}\Bigg).
\ee We may call (\ref{IRUVC1}) and (\ref{IRUVC2})  the ``classical"
IR-UV transition  because we are using Friedmann equations.

In order to generate chiral gravitational waves,  we consider the de
Sitter inflation by introducing a positive cosmological constant
$\bar{\Lambda}$ as
 \be
H^2=\frac{1}{2}\Big(\bar{\Lambda}-|\Lambda_W|\Big), \ee where we
choose $\bar{k}=0$ and $\bar{\Lambda}>|\Lambda_W|$. This leads to
the Sitter inflation like  $a(t)\sim e^{cHt}$. Then, we introduce
tensor perturbations only around de Sitter inflation~\cite{TS} \be
ds^2_{tp}=-dt^2+a^2(t)\Big[\delta_{ij}+h_{ij}(t,x_i)\Big] dx^i d
x^j,\ee where $h_{ij}$ is  a transverse-traceless tensor. At this
stage, we note that we choose $N^2=1$ for tensor perturbations, but
not $N^2=c^2$ as in Eq.(\ref{uvhamil1}) because  the former choice
makes the cosmological perturbation transparent.

Substituting this metric into the total action with $\alpha=\mu
c$, we find the bilinear action for $h_{ij}$ as \bea
\label{bilinear1}\delta^2S= \int dt d^3x a^3\Bigg[
\frac{1}{2\kappa^2}\dot{h}^i_j\dot{h}^j_i
-\frac{\kappa^2\mu^2\Lambda_W}{64a^2} h^i_j \bigtriangleup h^j_i
+\frac{\alpha}{4a^3}\epsilon^{ijk}h_{il} \bigtriangleup
h^l_{k|j}-\frac{\kappa^2\mu^2}{32a^4} h^i_j
\bigtriangleup^2 h^j_i\Bigg]\\
\label{bilinear2}=\int dt
d^3x\frac{a^3c^2}{2\kappa^2}\Bigg[\frac{\partial h^i_j}{\partial
x^0}\frac{\partial h^j_i}{\partial x^0}+ \frac{h^i_j \bigtriangleup
h^j_i}{a^2}+\frac{2}{a^3}\sqrt{-\frac{2}{\Lambda_W}}\epsilon^{ijk}h_{il}
\bigtriangleup h^l_{k|j}+\frac{2}{\Lambda_Wa^4} h^i_j
\bigtriangleup^2 h^j_i\Bigg], \eea where $\bigtriangleup$ denotes
the spatial  Laplacian and $x^0=c t$ with $[x^0]=-1$.

Tensor  $h_{ij}$ could be expanded in terms of plane waves with wave
vector $k_i$ with comoving momentum $k=\sqrt{k_ik^i}$ and $[k^2]=2$
as \be h_{ij}(t,x_i)= \sum_{A=L,R}\int\frac{d^3x}{(2\pi)^3}
\psi^A_{k}(t)e^{ix_ik^i}p^A_{ij}, \ee where $p^A_{ij}$ is a
circularly polarization tensor  defined by
$ik_s\epsilon^{rsj}p^A_{ij}=k\rho_A p^{Ar}~_i$.
$\rho^L=1(\rho^R=-1)$ denote for left (right)-handed circularly
polarized modes, respectively. Using the variable $v_{k}^A=a
\psi_{k}^A$ and the conformal time $\eta$ defined by $d\eta/dt=1/a$,
we obtain the Mukahnov-type equation for describing two circularly
polarized modes \be \frac{d^2v_{k}^A}{d\eta^2}+ \Big[(k^A_{\rm
eff})^2-\frac{2}{\eta^2}\Big]v_{k}^A=0,\ee where \be
\label{keff}(k^A_{\rm eff})^2=c^2a^2
\frac{k^2}{a^2}\Bigg(1-\rho^A\sqrt{-\frac{2}{\Lambda_W}}
\frac{k}{a}\Bigg)^2\ee with scaling dimensions $[\eta]=-2$ and
$[(k^A_{\rm eff})^2]=4$. On the other hand, $(k^A_{\rm eff})^2$ is
given  for $z=3$ HL gravity~\cite{TS} \be (k^A_{\rm eff})^2=c^2
a^2\frac{k^2}{a^2}\Bigg[1-\frac{2}{\Lambda_W}\frac{k^2}{a^2}\Big(1-\frac{2\rho^A}{\xi^2\mu}
\frac{k}{a}\Big)^2\Bigg].\ee At this stage, we wish to point out
that for $\alpha=\mu c$, two ghost-free circularly polarized waves
are generated. In the case of $\alpha\not=\mu c$, $(k^A_{\rm
eff})^2>0$ is not guaranteed, which may induce a ghost instability.
In the absence of gCS term, we find the ``quantum" IR-UV transition
as \be \label{IRUVq1}\frac{k^2}{a^2}\to
\frac{k^2}{a^2}\Bigg[1-\frac{2}{\Lambda_W}\frac{k^2}{a^2}\Bigg]. \ee
Expressing this  in terms of physical momentum
$p=k/a$ leads to \be \label{IRUVq3}p^2\to
p^2\Bigg[1-\frac{2}{\Lambda_W}p^2\Bigg]. \ee

\section{Circularly
polarized gravitational waves} In order to investigate tensor
propagations in the Minkowski background, we use the perturbation of
(\ref{kineticl})$+$(\ref{action4}). Equivalently, we make
substitution of $a\to 1$, $h_{ij}\to t_{ij}$, and $-\/\Lambda_W \to
\omega$ in Eq.(\ref{bilinear2}) to derive tensor perturbation. Then,
the field equation for tensor modes is given
by~\cite{My,Myung3,Myung4} \be \ddot{t}_{ij}-c^2 \bigtriangleup
t_{ij}
+\frac{2c^2}{\omega}\bigtriangleup^2t_{ij}-2c^2\sqrt{\frac{2}{\omega}}
\epsilon_{ilm}\partial^l \bigtriangleup t_j~^m=T_{ij} \ee with
 linearized-Cotton tensor $\delta C_{ij}=-\epsilon_{ilm}\partial^l\bigtriangleup t_j~^m$ and external
source $T_{ij}$. In this case,  we could not obtain
 the Euclidean covariant propagator because of the presence of Cotton tensor.
  Assuming a massless graviton propagation along the
$x^3$-direction with $p_{i}=(0,0,p_3)$, then  $t_{ij}(x^3)$ can be
expressed in terms of polarization components as~\cite{BS,APS,LQR}
\be
t_{ij}=\left(%
\begin{array}{ccc}
  t_+ & t_\times & 0 \\
  t_\times & -t_+ & 0 \\
  0 &0  & 0 \\
\end{array}%
\right). \ee Here,  the spatial  Laplacian  $\bigtriangleup$ reduces
to $\partial^2_3 ~(-p_3^2)$. Using this parametrization, we find two
coupled equations for different polarizations \bea \ddot{t}_+ -c^2
\bigtriangleup t_+
+2c^2\sqrt{\frac{2}{\omega}}\partial_3\bigtriangleup
t_\times+\frac{2c^2}{\omega}\bigtriangleup^2t_+=T_{+},
\\
\ddot{t}_\times -c^2 \bigtriangleup t_\times
-2c^2\sqrt{\frac{2}{\omega}}\partial_3\bigtriangleup t_+
+\frac{2c^2}{\omega}\bigtriangleup^2t_\times =T_{\times}. \eea To
find two independent components, we have to  introduce the
left-right base defined by \be t_{L/R}=\frac{1}{\sqrt{2}}\Big(t_+\pm
it_\times\Big) \ee where $t_L(t_R)$ represent the left
(right)-handed modes.  After Fourier-transformation, we find two
decoupled equations \bea \label{lhtm}-\varpi^2{t}_L+ c^2p_3^2 t_L
-2c^2\sqrt{\frac{2}{\omega}}p_3^3t_L+\frac{2c^2}{\omega}p_3^4t_L =
T_L,
\\
\label{rhtm}-\varpi^2{t}_R +c^2p_3^2 t_R
+2c^2\sqrt{\frac{2}{\omega}}p_3^3t_R+\frac{2c^2}{\omega}p_3^4t_R =
T_R \eea  with scaling dimension $[\varpi]=2$.

Finally, we have tensor propagators
 \be
\label{tenprpgCS}t_{L/R}=-\frac{T_{L/R}}{ \varpi^2-c^2p_3^2\Bigg(1
\mp\sqrt{\frac{2}{\omega}}p_3\Bigg)^2}\ee which shows clearly that
there is no ghost for two circularly polarized  modes $t_{L/R}$. It
is noted  from Eq.(\ref{lhtm}) that for $\alpha\not=\mu c$, one
could not make all spacial momentum terms  positive definite,
implying that a ghost state appears  for the  left-handed mode. Hence, the choice of
$\alpha=\mu c$ is essential  to obtain two ghost-free circularly
polarized waves.

In the absence of gCS term, the tensor propagators  take  the same
form \be \label{tenprp}t_{L/R}=-\frac{T_{L/R}}{
\varpi^2-c^2p^2\Big[1 +\frac{2}{\omega}p^2\Big]},\ee where  the
``quantum" IR-UV transition is observed as \be \label{IRUVq2}
 p^2 \to p^2\Big[1+\frac{2}{\omega} p^2\Big] \ee
with $p^2=p_ip^i$.

\section{GUP-corrected propagators}
 The GUP satisfies the
modified Heisenberg algebra~\cite{KMM,CMOK,KLM2}
 \bea
&&[x_i,p_j]=i\hbar \Big(\delta_{ij}+\beta p^2 \delta_{ij}+\beta' p_i p_j\Big),\nn \\
\label{UVCR}&&[x_i,x_j]=i\hbar \frac{(2\beta-\beta')+(2\beta+\beta')\beta p^2}{1+\beta p^2} \Big(p_i x_j-p_j x_i\Big),  \\
&&[p_i,p_j]=0, \nn \eea where $p_i$ is considered as the momentum at
high energies and thus, (\ref{UVCR}) can be interpreted to be the
UV-commutation relations. Here, we choose $\beta'=2\beta$ for
achieving the commutativity with UV scaling dimension $[\beta]=-2$. In this case,
the minimal length
 which follows from the modified Heisenberg algebra is given by
 \be
 \delta x_{\rm min}=\hbar \sqrt{5\beta}.\ee
 The presence of the minimal length represents  a feature of the GUP.
On the other hand, introducing IR-canonical variable  $p_{0i}$ with
$x_i=x_{0i}$ through the replacement \be \label{RPCR}p_i \to
p_{0i}\Big(1+\beta p_0^2\Big),\ee these variables satisfy canonical
commutation relations \be \label{IRCR}
[x_{0i},p_{0j}]=i\hbar\delta_{ij},~[x_{0i},x_{0j}]=[p_{0i},p_{0j}]=0.
\ee Here  $p_{0i}$ is considered as the momentum at low energies (IR
region) with $p_0^2=p_{0i}p_{0i}$. It is easy to show that
Eq.(\ref{UVCR}) is satisfied to $\beta$-order when using Eq.
(\ref{IRCR}). Hence, the replacement (\ref{RPCR}) could be used as the
``low-energy window" to see quantum gravity phenomenology up to
$\beta$-order. However,  at $\beta^2$-order, we observe that the
translation invariance is broken because coordinate $x_i$ becomes
noncommutative. Thus, it is not guaranteed that the replacement
(\ref{RPCR}) works up to $\beta^2$-order. Accordingly, the following
replacement is suggested  for making a connection between GUP and
$z=2$ HL gravity \be \label{RPCRn}  p^2 \to p_0^2\Big(1+2\beta
p_0^2\Big).\ee

At high energies, we assume that the UV-propagator has   \be
\label{UVprop} G_{\rm UV}(\varpi,p^2)=\frac{1}{\varpi^2-c^2p^2},\ee
whereas at low energies,  the IR-propagator takes the conventional
form  \be \label{IRprop}G_{\rm
IR}(\varpi,p^2_{0})=\frac{1}{\varpi^2-c^2p_{0}^2}.\ee Making use of
(\ref{RPCRn}), the UV-propagator (\ref{UVprop})  can be rewritten as
\be G_{\rm UV}(\varpi,p_{0}^2)=\frac{1}{\varpi^2-c^2p_0^2\Big(1+
2\beta p_{0}^2\Big)}. \ee The GUP-corrected tensor propagator is
determined by  \be \label{gupcoprop} t^{GUP}_{ij} = -G_{\rm
UV}(\varpi,p_0^2)T_{ij}=-\frac{T_{ij}}{\varpi^2-c^2p_0^2\Big(
1+2\beta p_{0}^2\Big)}. \ee  This is  the same form as the UV-tensor
propagator (\ref{tenprp}) when using the replacement of $\beta\to
1/\omega$ and $p^2 \to p^2_0$ for the $z=2$ HL gravity.

Now we may  make a connection between GUP and $z=2$ HL gravity with
gCS term  (circular polarized tensor modes). To this end, we
introduce doubly special relativity which suggests  modification of
commutators~\cite{ADV} \be \label{newcr}
[x_i,p_j]=i\hbar\Big[\delta_{ij}-\sqrt{\beta}\Big(p\delta_{ij}+\frac{p_ip_j}{p^2}
\Big)+\beta\Big(p^2\delta_{ij}+3p_ip_j\Big)\Big] \ee with
$[x_i,x_j]=[p_i,p_j]=0$. Then, let us define \be \label{newIR}
x_i=x_{0i},~~p_i \to p_{0i}\Big(1-\sqrt{\beta}p_0+2\beta
p_0^2\Big),\ee where $x_{0i}$ and $p_{0j}$ satisfy the canonical
commutation relations $[x_{0i},p_{0j}]=i\hbar \delta_{ij}$. It could
be shown that Eq.(\ref{newcr}) is satisfied at $\beta$-order. From
Eq.(\ref{newIR}),  we have \be p^2 \to
p_0^2\Big(1-2\sqrt{\beta}p_0+4\beta p_0^2\Big)\ee up to
$\beta$-order. However, this is not the case of left-handed  mode in
Eq.(\ref{tenprpgCS}). Hence, the connection between GUP and
circularly polarized tensor modes is not clearly defined.

\section{Discussions}

We have constructed the chiral gravitational waves from the $z=2$
Ho\v{r}ava-Lifshitz gravity with gCS term in the de Sitter and
Minkowski backgrounds, as the $z=3$ Ho\v{r}ava-Lifshitz gravity did
provide them~\cite{TS}. It turns out that the  gCS term plays an
essential role in making circularly polarized gravitational waves in
the $z=2$ and $z=3$ Ho\v{r}ava-Lifshitz gravity theories.
Particularly, its coefficient should be chosen as $\alpha=\mu c$ to
obtain ghost-free chiral gravitational waves.

Also, we observe   classical and quantum IR-UV transition rules
which show the feature of the $z=2$ HL gravity: (\ref{IRUVC1}) and
(\ref{IRUVC2}) for classical rules and (\ref{IRUVq3}) and
(\ref{IRUVq2}) for quantum rules. As is shown in Eq.(\ref{RPCRn}),
quantum transition rule may be explained from the GUP  when
replacing $-\frac{1}{\Lambda_W}$ and $\frac{1}{\omega}$ by $\beta$.
The modified Heisenberg commutation (\ref{UVCR}) is satisfied to
$\beta$-order. Thus, the replacement (\ref{RPCRn}) was used to
derive the GUP-corrected propagator (\ref{gupcoprop}) which is the
same form as the UV-tensor propagator (\ref{tenprp}). Hence, the
$z=2$ deformed HL gravity without gCS term is well interpreted by
the GUP.  We note   that $-\frac{2}{\Lambda_W}\frac{k^4}{a^4}$ in
the Mukahnov-type equation (\ref{keff}) and
$\frac{2}{\omega}p^4$-term in the tensor propagator (\ref{tenprp})
come solely from $R^2_{ij}$-term in Eqs. (\ref{action3}) and
(\ref{action4}), respectively. Therefore, $R_{ij}^2$-term is
responsible for showing the quantum aspects of the $z=2$ HL gravity.

On the other hand, classical IR-UV transition rules remain unchanged
in the presence of gCS term. However, quantum transition rules do
not work. Even though a modified commutation (\ref{newcr}) was
introduced, the connection between circularly polarized tensor modes
and GUP is not clearly established.

In conclusion, the $z=2$ HL gravity may be considered as a candidate
for the quantum Einstein gravity. It provides matter bounce solution
and chiral gravitational waves, as obtained from the $z=3$ HL
gravity. It was shown that the renormalized Wheeler-DeWitt equation
possesses a solution with a $z=2$ Lifshitz point, but no other $z>2$
solutions to leading order of strong coupling expansion~\cite{Sak}.
Considering  the $z=2$ HL gravity with the gCS term,  their
conclusion remains unchanged.  Moreover, the perturbative
corrections to $z=2$ Lifshitz point are more attractive than the
$z=3$ Lifshitz point.

\section*{Acknowledgement}
This work  was supported by Basic Science Research Program through
the National Research  Foundation (KRF)  of  Korea funded by the
Ministry of Education, Science and Technology (2009-0086861).

\end{document}